# The Structure of the Planetary Nebula NGC 2371 in the Visible and Mid-Infrared

Gerardo Ramos-Larios, J.P. Phillips[†]

Instituto de Astronomía y Meteorología, Av. Vallarta No. 2602, Col. Arcos Vallarta, C.P. 44130 Guadalajara, Jalisco, México   e-mail: gerardo@astro.iam.udg.mx

[†]  Deceased, 2011 April 29.

**Abstract**

We investigate the structure of the planetary nebula (PN) NGC 2371 using [OIII] $\lambda5007$ imaging taken with the Jacobus Kapteyn 1.0 m telescope, and [NII] $\lambda6584$, [OIII] $\lambda5007$ and H$\alpha$ results acquired with the *Hubble Space Telescope* (*HST*). These are supplemented with archival mid-infrared (MIR) observations taken with the *Spitzer Space Telescope* (*Spitzer*). We note the presence of off-axis low-ionization spokes along a PA of 65°, and associated collars of enhanced [OIII] emission. The spokes appear to consist of dense condensations having low-excitation tails, possibly arising due to UV shadowing and/or ram-pressure stripping of material. Line ratios imply that most of the emission arises through photo-ionisation, and is unlikely to derive from post-shock cooling regions. An analysis of these features in the MIR suggests that they may also be associated with high levels of emission from polycyclic aromatic hydrocarbons (PAHs), together with various permitted and forbidden line transitions. Such high levels of PAH emission, where they are confirmed, may develop as a result of preferentially enhanced FUV pumping of the molecules, or shattering of larger grains within local shocks. Although H$_2$ emission may also contribute to these trends, it is argued that shock-excited transitions would lead to markedly differing results. We finally note that thin filaments and ridges of [OIII] emission may indicate the presence of shock activity at the limits of the interior envelope, as well as at various positions within the shell itself. We also note that radially increasing fluxes at 3.6, 5.8 and 8.0 $\mu$m, relative to the emission at 4.5 $\mu$m, may arise due to enhanced PAH emission in external photo-dissociative regions (PDRs).

**Key Words:** planetary nebulae: individual: NGC 2371 --- ISM: jets and outflows --- infrared: ISM --- ISM: lines and bands

# 1. Introduction

The planetary nebula (PN) NGC 2371 has what appears to be a regular elliptical shell, divided into two main segments by a dark major-axis lane. This is surmounted by symmetrically placed ansae located ± 66 arcsec from the central star (e.g. Sabbadin et al. 1982; Gorny et al. 1999), which have recently been found to represent the external limits of bilobal outflow structures (Ayala et al. 2005). The estimated He II Zanstra temperature of the source ($T_Z(HeII) \cong 99.6$ kK; see e.g. Phillips (2003) for a summary of prior estimates of this parameter) is significantly larger than for $T_Z(HI)$ ($\cong 49.7$ kK), a difference which is usually interpreted as indicating the escape of H ionizing photons from the nebular shell. This, together with the absence of appreciable excess emission in the infrared (IR) (e.g. Pottasch et al. 1981); the failure to detect CO or $H_2$ transitions (Huggins et al. 1996; Kastner et al. 1996); and the low levels of combined interstellar and nebular extinction (e.g. Tylenda et al. 1992; Aller & Czyzak 1979; Cahn 1976; Torres-Peimbert & Peimbert 1997; Herald & Bianchi 2004) suggest that the shell is density bound, and HI/HII mass ratios are relatively small.

One of the interesting features in the source is the presence of low-ionization spokes or jet-like structures along a PA of 65°, and associated collars of enhanced [OIII] emission. The spokes appear to consist of dense condensations having low-excitation tails, possibly arising due to UV shadowing and/or ram-pressure stripping of material that are also visible in the MIR. These structures would be genuine jets, but the lack of kinematical data of the source does not allow us to determine the true nature of the spokes which deserve a more detailed investigation than has been available heretofore.
In a pair of excellent works (Gonçalves et al. 2001, 2004), found similar structures in around 50 PNe, classified in terms of their morphology and kinematics. Also, explain that the knots that appear in symmetrical and opposite pairs of low velocity could be understood as the survival of fossil condensations formed during the AGB phase or probably structures that have experienced substantial slowing down by the medium.

We shall analyse a broad range of archival material relating to this source, permitting us to reveal facets of the structure which have not been noted. Thus, we will use narrow band data deriving from the *Hubble Space Telescope* (*HST*) to show the presence of low-ionization spokes – hereafter referred to as LISp, in the outer portions of the shell, and employ line ratio diagnostics to investigate the natures of these regions. Finally, we shall study the nature of NGC 2371 in the mid-infrared (MIR), using images deriving from the *Spitzer Space Telescope* (*Spitzer*). These will be used to investigate the infrared structure in four photometric bands, and to define the MIR properties of the narrow LISp features.

## 2. Observations

### 2.1 Narrow Band Imaging with the HST

The *HST* observations[1] of NGC 2371 were obtained on 2007/11/15 using the Wide Field Planetary Camera 2 (WFPC2; Holtzman et al. 1995; see also the WFPC2 Instrument Handbook (Biretta 1996)) with aperture WF3-FIX, and form part of *HST* program 11093 (Hubble Heritage Observations of PNe with WFPC2; principal investigator (P.I.): Keith Noll). The enhanced imaging results were downloaded from the Hubble Legacy Archive at http://hla.stsci.edu/, are astrometrically corrected, and in units of electrons/sec. We have selected exposures taken with filter F502N (pivot wavelength $\lambda_p$ = 5012.4 Å, effective bandwidth $\Delta\lambda$ = 26.9 Å) corresponding to emission from the $\lambda$5007 transition of [OIII], and a total exposure time of $\Delta t$ = 1600 s; filter F658N ($\lambda_p$ = 6590.8 Å, $\Delta\lambda$ = 28.5 Å, $\Delta t$ = 1600 s, [NII] $\lambda$6584 Å); filter F656N ($\lambda_C$ = 6563.8 Å, $\Delta\lambda$ = 21.5 Å, $\Delta t$ = 1600 s, H$\alpha$ $\lambda$6563 Å); and filter 673N ($\lambda_C$ = 6732.2 Å, $\Delta\lambda$ = 47.2 Å, $\Delta t$ = 1600 s, [SII] $\lambda$6717+6732 Å). The sizes of the pixels in the images is 0.1 arcsec, and the point source spread function (PSF), as measured by profiles through stars, has a FWHM of 0.2 arcsec.

---

[1] Based on observations made with the NASA/ESA Hubble Space Telescope, and obtained from the Hubble Legacy Archive, which is a collaboration between the Space Telescope Science Institute (STScI/NASA), the Space Telescope European Coordinating Facility (ST-ECF/ESA) and the Canadian Astronomy Data Centre (CADC/NRC/CSA).

## 2.2 Spitzer Imaging in the MIR

The *Spitzer*[2] imaging of NGC 2371 with the Infrared Array Camera (IRAC; Fazio et al. 2004) was taken on 27/11/2006 as part of Program 30285 (*Spitzer* Observations of Planetary Nebulae-2; P.I. Giovanni Fazio), and the results have been processed as described in the IRAC Instrument Handbook Version 1.0, February 2010 (available at http://ssc.spitzer.caltech.edu/irac/iracinstrumenthandbook/IRAC_Instrument_Handbook.pdf). The resulting post-Basic Calibrated Data (post-BCD) are relatively free from artefacts; well calibrated in units of MJy sr$^{-1}$; and have reasonably flat emission backgrounds. The observations employed filters having isophotal wavelengths (and bandwidths $\Delta\lambda$) of 3.550 $\mu$m ($\Delta\lambda$ = 0.75 $\mu$m), 4.493 $\mu$m ($\Delta\lambda$ = 1.9015 $\mu$m), 5.731 $\mu$m ($\Delta\lambda$ = 1.425 $\mu$m) and 7.872 $\mu$m ($\Delta\lambda$ = 2.905 $\mu$m). The normal spatial resolution for this instrument varies between ~1.7 and ~2 arcsec (Fazio et al. 2004).

## 2.3 Narrow Band Imaging with the JKT

Narrow band [OIII] observations of NGC 2371 were also acquired on 8/10/92 using the 1.0 m JKT, based in La Palma, Spain. They are taken from the publicly available Isaac Newton Group (ING) archive[3], accessible at http://casu.ast.cam.ac.uk/casuadc/archives/ingarch. The CCD-EEV7 detector had a pixel size of 22 microns, and was mounted at the f/15 Cassegrain focus of the telescope. Given that the plate scale of the JKT is 13.8 seconds of arcsec/mm, each pixel subtends an angle of 0.33×0.33 arcsec on the sky. Finally, the air mass was 1.40, and exposure time was 1000 sec, whilst profiles through stars in the field indicates combined seeing/guiding sizes of 0.95 arcsec.

## 3. The Visual Morphology of NGC 2371

A deep [OIII] image of NGC 2371 taken with the JKT is illustrated in the lower panel of Fig. 1, where the results have been processed using unsharp masking techniques. We have additionally used the IRAF[4] task "crmedian" to remove cosmic rays in pixels deviating by a specified statistical amount from the median at each pixel. Finally, we have and applied a Low-Pass Kernel filter using a commercial program MAXIM DL produced by Diffraction Limited.

---

[2] This work is based, in part, on observations made with the Spitzer Space Telescope, which is operated by the Jet Propulsion Laboratory, California Institute of Technology under a contract with NASA.

[3] This paper makes use of data obtained from the Isaac Newton Group Archive which is maintained as part of the CASU Astronomical Data Centre at the Institute of Astronomy, Cambridge.

It is clear from this that the limits of the elliptical envelope show evidence for bright unresolved rims similar to those previously noted in other PNe (e.g. Phillips et al. 2010; Balick 2004; Medina et al. 2009). Such features have been attributed to shock interaction between differing elements of the nebular shells, leading to enhanced levels of emission within post-shock cooling regions.

Even higher resolution observations of the central shell are illustrated in Fig. 2, where we present *HST* images of the source in [OIII] $\lambda$5007 (blue), H$\alpha$ (green) and [NII] $\lambda$6584 (red) (left-hand panel), and ratios between [OIII] and H$\alpha$ (in the right-hand panel). The images have again been processed using unsharp masking techniques. Several interesting aspects of the structure are evident from a casual inspection of these results, including narrow ridges of [OIII]/H$\alpha$ enhancement at the limits of the primary shell, noted previously in our discussion of the JKT results (Fig. 1). Several barely resolved condensations, many of which have tails extending to ~ 0.8 arcsec, appear to be particularly bright in the low excitation transitions (left-hand panel). It is possible that these are similar to the tadpole-type structures observed in NGC 7293 (the "Helix"), attributed to shadowing effects from dense clumps (Cantó et al. 1998), flows overrunning a slowly expanding system of pre-existing dense neutral globules (Dyson et al. 2006) or produced from larger clouds through break up due to Rayleigh-Taylor instability (Capriotti & Kendall 2006).

Finally, we note the presence of two LISp features oriented along a PA of ~ 65°, commencing at distances ~16.5 arcsec from the position of the central star. Both of these are associated with crescent-shaped collars of enhanced [OIII] emission, evidence for which is also present in the JKT results. The edges of these collars appear to be surprisingly sharply defined, and will be discussed in further detail in our analysis in

---

[4] IRAF, the Image Reduction and Analysis Facility, is distributed by the National Optical Astronomy Observatory, which is operated by the Association of Universities for Research in Astronomy (AURA) under cooperative agreement with the National Science Foundation

Sect. 5.

These various optical components are also seen in the profiles in Fig. 3, where we show the variation of line intensities as a function of distance from the nucleus. It will be noted that there is a very sharp decline in line strengths at the position of the [OIII] collars, for instance, located at radial positions (RPs) of ~14 arcsec from the central star. This also leads to some change in [OIII]/H$\alpha$ ratios, and lesser variations in the lower excitation lines (Fig. 4).

Large variations in line strengths and ratios are also noted for the LISp, for which we find increases in [NII]/H$\alpha$ and [SII]/H$\alpha$ at radial distances > 16 arcsec, and a corresponding decline in [OIII]/H$\alpha$ (Fig. 4). The properties of these structures will be investigated later in this analysis (Sect. 6).

## 4. The Structure of NGC 2371 in the Mid-Infrared

A *Spitzer* MIR image of NGC 2371 is illustrated in Fig. 1, where we have combined results from three of the IRAC channels. Emission at 3.6 $\mu$m is represented as blue, 4.5 $\mu$m by green, and 8.0 $\mu$m by red. It is apparent that the morphology of the source is similar to that noted in the visible, and includes weak evidence for lobes along the major axis, and LISp along PA = 65°.

Major axis profiles through the symmetry axes of the lobes are illustrated in Fig. 5. This involved the removal of large background offsets, as well as much smaller linear gradients. We have also obtained profiles of the 8.0$\mu$m/4.5$\mu$m, 5.8$\mu$m/4.5$\mu$m and 3.6$\mu$m/4.5$\mu$m flux ratios. Such ratios tend to show the distribution of PAHs since there is very little emission by polycyclic aromatic hydrocarbons (PAHs) within the 4.5 $\mu$m channel.

Some care must be taken in interpreting the latter results, since scattering within the IRAC camera can lead to errors in relative intensities. The flux corrections for extended source photometry lead to maximum changes of ~0.91 at 3.6 $\mu$m, 0.94 at 4.5 $\mu$m, 0.66-0.73 at 5.8 $\mu$m and 0.74 at 8.0 $\mu$m (IRAC Instrument Handbook), although the precise values of these corrections also depends on the distribution of surface brightness in the source.

Several aspects of these profiles are of particular interest. It is evident that there is a marked dip in emission at the centre of the source (i.e. for RPs of -9→5 arcsec), and outerlying components of emission at RP $\cong$ ±60 arcsec. The latter features, designated as the SE and NW ansae, are located at the outer limits of the visual bilobal structure. It is also apparent that the highest levels emission are associated with the 4.5 $\mu$m channel, where fluxes are ~ 7 times greater than at 3.6 $\mu$m. This trend is quite different from what has been observed in most other PNe, where 3.6$\mu$m/4.5$\mu$m ratios are usually closer to unity (see e.g. Phillips & Ramos-Larios 2008a; Ramos-Larios & Phillips 2008). We also see an increase in 8.0$\mu$m/4.5$\mu$m and 5.8$\mu$m/4.5$\mu$m ratios with distance from the central star, a variation which has been noted in many other PNe (Phillips & Ramos-Larios 2008a; Ramos-Larios & Phillips 2008) and HII regions (e.g. Phillips & Ramos-Larios 2008b; Phillips & Perez-Grana 2009). Given the high gas-phase C/O ratios in NGC 2371 (~2; e.g. Milanova & Kholtygin 2009; Peimbert & Torres-Peimbert 1983), this may suggest increasing levels of PAH emission towards the periphery of the source, located either within neutral condensations and/or exterior post-dissociative regions (PDRs).

These trends are difficult to follow out to very much greater distances given the decreasing levels of nebular emission, although we note that mean ratios for the NW and SE ansae appear to be greater still, of respective orders 5.8$\mu$m/4.5$\mu$m ~1 & 1.7, and 8.0$\mu$m/4.5$\mu$m ~2.2 & 2.7.

Trends along an axis including the two LISp features are illustrated in Figs. 3 and 4, where they may be directly compared to the *HST* results. The differing spatial resolutions of the Spitzer (~ 2 arcsec) and *HST* (~0.2 arcsec) observations leads to a relative blurring of the trends, although it is clear that several of the features are present in both of these sets of profiles. The dramatic increase in lower excitation line strengths at the position of the SW LISp (Fig. 3, RP > 15 arcsec) is present in most of the IRAC results – although there is very little evidence for corresponding emission in the 4.5 $\mu$m channel; a trend which will be used to investigate the nature of the MIR emission in Sect. 6.2. By contrast, the sharp decline in ionic line strengths at the position of the SW shock, mentioned in our discussion of the *HST* results in Sect. 3, appears not to be present in any of the IRAC bands – although the limited resolution of the MIR results, and the strength of the LISp complicates the interpretation of trends in this particular region.

Finally, the variation in the IRAC band ratios is illustrated in the lower panel of Fig. 4, where the SW LISp feature is seen to lead to significant deviations in the profile trends. There may also be evidence for a weak enhancement in ratios for the NE LISp as well. By contrast, there is again little evidence for variations in profile due to the [OIII] collar/shock regions.

IRAC band ratio mapping is illustrated in Fig. 6, where it can be seen that the central portions of the source possess lower flux ratios, with minima occurring ~ 7 arcsec from the central star. The ratios subsequently increase to larger radial distances, as previously noted in our profiles in Figs. 4 & 5, with the largest values occurring at the periphery of the shell. The greater complexity of structure in the 5.8$\mu$m/4.5$\mu$m map is likely to be a consequence of the lower S/N of the 5.8 $\mu$m results.

## 5. The Nature of the SW Region

The collars of sharply defined [OIII] emission just inside the LISp features have all of the appearance of being associated with the LISp, and causatively linked to their formation, and it is likely that an understanding of these structures will help in defining the LISp formation process itself.

There have, in this context, been several attempts to model such LISp. Thus for instance, it has been suggested that shock refraction of stellar winds leads to focussing of material, and the formation of narrowly defined jets at the major axis limits of the nebular envelopes (e.g. Frank et al. 1996; Canto et al. 1988). This may explain the presence of ansae, jets and FLIERS in sources such as NGC 3242, NGC 6543, NGC 6826, and NGC 7009 (see e.g. Balick et al. 1994, 1998), but the Canto model will not apply for highly numerous knots and tails in many orientations. It is far from clear how such analyses would be consistent with the present source, however, where the LISp occur along a PA ~ 65° which differs from that of the nebular major and minor axis. A rather better explanation may therefore be afforded by the analysis of García-Segura (1997), where it is suggested that magnetic collimation around a precessing star may lead to the formation of off-axis jets.
Very similar structures have been reported in K1-2 and Wray 17-1 (Corradi et al. 1999). Their kinematical study shows an increasing velocity outward as a clump, suggesting a density stratification of the ambient gas for the first case and an effect of ionization for Wray 17-1 (Gonçalves et al. 2001).

Another explanation is proposed by Steffen & Lopez (2004), who point out that ablation flows have lower densities or smaller sizes than their source clouds, resulting in higher accelerations and terminal velocities that approach to the medium. And because the initial acceleration for low-density gas is high, the ablation tails of high-density are seen as the velocity spikes.

None of these models predicts the [OIII] collars observed in NGC 2371, however – although it is tempting to suggest that the narrowly collimated flows are punching through the primary envelope of the source, and leading to an annular void of lower intensity emission. It would be plausible, under such circumstances, to presume that the sharply defined crescent rims arise through local shock activity.

Mellema et al. (1998) have studied the evolution of dense clumps being gradually photo-ionised by the stellar radiation. At the ionization front, material is ionised and thrown out in a photo-evaporation flow. When the shock has moved through the clump, begins the cometary phase in which the clump accelerate caused by the photo-evaporation flow. It therefore seems that a reliable explanation represents a challenge for any mechanism purporting to explain this phenomenon

A profile through the SW collar is illustrated in Fig. 7, where we show the variations in the ratios [NII]/H$\alpha$, [SII]/H$\alpha$ and [OIII]/H$\alpha$. The ratios have been normalised to unity at RP = 0 arcsec. Where shocks are of relevance, and the propagation vectors are inclined to the line of sight, then one might expect that low excitation ratios would increase in the post-shock region, and lead to peaks close to (but slightly displaced from) the frontal position. It is clear from Fig. 7 that this does not seem to be occurring, however, and there is very little deviation in either of these profiles. By contrast, it is evident that values of [OIII]/H$\alpha$ decline by ~ 20 % within 0.3 arcsec of the putative frontal position.

This inconsistency between shock modelling and profiles doesn't necessarily imply that shocks are irrelevant to this region. It may be possible for instance that Mach numbers are high, and frontal temperatures and UV fluxes are appreciable, leading to local ionization of the post- and pre-shock gas. This, together with irradiation by the CS radiation field, may be sufficient to suppress the levels of lower excitation emission.

Where shocks are not, on the other hand, directly responsible for this feature, then one must explain the extreme sharpness with which these collars are defined. A possible and simply explanation is that the [OIII] collars are just thin transition regions, where the ionisation of O drops from the $O^{+++}$ in the nebular interior to $O^+$ in the rim.

We have noted that the primary fall-off in [OIII] emission occurs within a distance of ~ 0.3 arcsec. Whilst the distance of NGC 2371 is not very well determined, statistical estimates imply that D = 1.32→2.07 kpc (e.g. Phillips 2004; Zhang 1995; van de Steene & Zijlstra 1994), and we shall adopt a mean value here of 1.7 kpc. This would imply that the fall-off in emission occurs over a distance ~ $2.5\ 10^{-3}$ pc.

Given that the most reliable estimates of electron temperature imply that $T_e \cong 1.5\ 10^4$ K (see e.g. Pottasch et al. 1981; Torres-Peimbert & Peimbert 1977; Natta et al. 1980; Olguín et al. 2002), then corresponding sound speeds would be of order ~ 20 km $s^{-1}$, and one would expect such edges to disperse over periods of ≈ 120 yrs. Where this collar is not formed and being maintained by ongoing shock activity, it therefore follows that the formation of the structures must have been very recent indeed.

## 6. Low-Ionization Spokes Emission Processes

### 6.1 Analysis of the *HST* Results

We have already noted that narrowly defined LISp, such as those in Fig. 2, may be formed though a variety of shock and/or magnetic processes, whilst the off-axis location of these features suggests precession of the central star.

The nature of the emission in these LISp is far from clear, however. It is apparent, from Fig. 8, that the LISp is composed of multiple narrow features, appearing to take the form of tail-like structures attached to barely resolved condensations. Profiles though the condensations imply a typical deconvolved width of ≤ 0.1 arcsec, and corresponding physical dimensions of order < $8.2\ 10^{-3}$ pc. The expansion timescales of < 200 yrs for $T_e = 1.5\ 10^4$ K would be very much less than the likely expansion period of the envelope as a whole (of order $5.5\ 10^3$ yrs where one takes an expansion velocity of 42.5 km s$^{-1}$ (Sabbadin 1992)). This would therefore suggest that we may be dealing with neutral condensations within or at the limits of the ionised envelopes, and that much of the emission derives from thin ionised skins about the denser HI cores. The high opacities of these regions would plausibly lead to UV shadows, the structures and characteristics of which were first described by Capriotti (1973) and Van Blerkom & Arny (1972). Under these circumstances, much of the ionisation occurs through irradiation by diffuse photons from the surrounding gas, and results in enhanced levels of lower excitation emission. It is also possible that the tails are partially formed through ram-pressure stripping of the condensations, however - a process which would lead to similar comet-like structures, and greatly limit the lifetimes of the condensations.

A profile taken at the base of the SW LISp is illustrated in Fig. 7, where it is clear that there is a marked increase in [SII]/H$\alpha$ and [NII]/H$\alpha$ ratios within a distance of ~ 0.2 arcsec, and a corresponding decrease in [OIII]/H$\alpha$. Much of this variation is associated with an individual clump. The low excitation ratios subsequently decline within the tail of the clump – although they remain in excess of values outside of the LISp (i.e. at RPs < 0 arcsec). Similarly, [OIII]/H$\alpha$ ratios increase within the tail, but are again less than the values observed outside of the LISp. Comparable trends are also evident in the montage of images presented in Fig. 8, where it is again apparent that [OIII] ratios are suppressed, and [NII] and [SII] ratios are enhanced.

All of these characteristics are qualitatively consistent with the hypotheses of neutral condensations, and it is tempting to suggest that the processes leading to the LISp have resulted in the disruption of an enveloping neutral region. It is unclear however whether the range of radii $\Delta r \approx 8$ arcsec over which the condensations are observed is representative of the width of the neutral shell, or arises through acceleration of individual clumps – whether this occurs through shock interaction within the LISp flow, or asymmetric heating and mass-loss from the clumps.

This is not the only explanation capable of explaining the present trends, however. We have already noted that there is very little evidence for neutral gas in NGC 2371 (Sect. 1), and this opens the possibility that other mechanisms may be required to explain such emission. We note for instance that the interaction of fast, collimated stellar winds with denser (but not necessarily neutral) shell irregularities may lead to the formation of post-shock regions with somewhat similar characteristics. In particular, the lengths of the tails appear to be ~ 0.8 arcsec, corresponding to ~ 2 $10^{16}$ cm for the distances cited above. Such dimensions are similar to those expected for low excitation emission within post-shock cooling regions. Thus, whilst the scale-length for cooling depends upon the physical characteristics of the shocks, including the pre-shock densities, shock velocities, and role of magnetic fields, we note that most models suggest typical values of a few times $10^{16}$ cm which are not dissimilar to the tail lengths observed in the present source (see e.g. Ohtani 1980; Pittard et al. 2005; Dopita 1977; Dopita & Sutherland, 1996; Shull & McKee 1979).

One way of assessing the importance of these mechanisms is through the diagnostic diagram illustrated in Fig. 9, where we have represented the logarithms of [SII]/H$\alpha$ against [OIII]/H$\alpha$ (upper panel), and of H$\alpha$/[NII] against H$\alpha$/[SII] (lower panel). Riera et al. (2008) found an overlap in the values of the [OIII]/H$\alpha$ ratios between photo-ionised gas and shocks. The integrated emission for shocks shows a big change in the [SII]/H$\alpha$ and [NII]/H$\alpha$ ratios with intermediate values expected for a pure shock-excited and photo-ionised nebula.The regions over which ratios have been determined are indicated in the inserted images, and includes the primary emission associated with the SW LISp. In the case of the upper panel, the blue diamonds correspond to individual measures of 492 PNe, using a heterogeneous sample of spectra deriving from Kaler et al. (1997). These are corrected for interstellar extinction using the coefficients of Tylenda et al. (1992), and the extinction curve of Savage & Mathis (1979). The large red disks, on the other hand, correspond to the shock models of Hartigan et al. (1987) and Shull & McKee (1979).
Finally, the present *HST* ratios results are illustrated using small green diamonds, where we have determined ratios on a pixel-by-pixel basis, and corrected for quantum throughputs QT, and the differing wavelengths of the lines. Values of QT are taken from the WFPC2 Instrument Handbook (Biretta 1996), where the procedure for transforming observational measures (in electrons $s^{-1}$ pixel$^{-1}$) to fluxes (in ergs cm$^{-2}$ s$^{-1}$ arcsec$^{-2}$) is also explained.

We have taken into account the leaking of [NII] $\lambda6548$ and $\lambda6584$ emission within the H$\alpha$ filter F656N, and of H$\alpha$ emission within the [NII] $\lambda6584$ filter F658N.

The system through-puts in the H$\alpha$ (F656N) filter are of the order of QT = 0.113 for H$\alpha$, 0.034 for [NII] $\lambda6548$, and 0.0066 for [NII] $\lambda6584$, and similar parameters also apply for the F658N filter (QT = 0.112 for [NII], and 0.0043 for H$\alpha$). Given the ratios [NII]/H$\alpha$ implied by our present results, this would imply that the H$\alpha$/[NII] and H$\alpha$/[SII] ratios are too high. It follows that the points in Fig. 9 need to be shifted downwards and to the left by ~0.09 dex, closer to the SNR zone.

In a comparison with the results of Raga et al. (2008), it is apparent that LISp ratios fall within a narrowly defined range of values, and are most closely associated with the ratios found in lower-excitation PNe. They are inconsistent with the ratios predicted though shock modelling. It would therefore appear that neutral condensations/UV shadow tails may contribute much of the emission in these LISp.

Further evidence for this is illustrated in the lower panel of Fig 9, where we have undertaken a similar analysis in the H$\alpha$/[NII]-H$\alpha$/[SII] colour plane. In order of detect thin shocks, the *HST* ratios have been also processed in a similar way to that described for the JKT imaging (see Section 3). The ranges of ratio for supernova remnants (SNRs) and HII regions are taken from Sabbadin, Minello, & Bianchini (1977), whilst the ratios for PNe are derived from Riesgo & Lopez (2006). Where shocks are important, then one might anticipate that ratios would be strongly biased towards the region of the SNRs. This is not, however, the case, and it is clear that the ratios are more consistent with those in (radiatively excited) Galactic PNe.

We therefore conclude that the LISp in NGC 2371 are characterized by low levels of excitation, with the emission deriving from photo-ionised regions. Much of the flux is also likely to be associated with neutral clumps, and their radially directed tails. However, the LISp are also evident in the IR as well, where emission processes may be markedly different. It is of interest to investigate the likely contributions to fluxes in this wavelength range as well.

## 6.2 Low-Ionization Spokes Emission Characteristics in the MIR

Although the spatial resolution of the *Spitzer* mapping is ~10 times less than that of the *HST*, we have already noted that the LISp in NGC 2371 are visible at these wavelengths as well (see Fig. 1). The features are seen to have significantly greater levels emission at 3.6, 5.8 and 8.0 $\mu$m than is the case in the surrounding envelope.

We note that ionic transitions, $H_2$ emission, dust continuum and PAH emission may all play a role in enhancing emission within the LISp, and that these have differing characteristics within the MIR colour plane. To assess what the intrinsic colours of the LISp are in the MIR, however, one must first isolate the LISp contribution from that of the underlying envelope.

This is not entirely easy to achieve, given that primary shell emission is relatively strong – and the present results will, as a result, be subject to a certain level of uncertainty. However, the characteristics of the SW LISp appears to be particularly helpful in this respect, and we have illustrated the trends for this feature in Fig. 10. Band strengths are normalized to unity at RP = 14.65 arcsec.

It is apparent that whilst the 4.5 $\mu$m fluxes show a monotonic decline between RP = 13 and 30 arcsec, with no indication of deviations due to the LISp, the other bands show increases in fluxes at the precise position of the LISp. We shall assume from this that whilst the LISp is strong in the 3.6, 5.8 and 8.0 $\mu$m channels, it is all but absent at 4.5 $\mu$m. This is used to determine intrinsic LISp flux ratios of order $3.6\mu m/5.8\mu m \cong 1.85$, and $8.0\mu m/5.8\mu m \cong 5.44$.
To gain a further insight into the possible nature of the emission, we have indicated the position of the LISp with the $3.6\mu m/5.8\mu m$-$8.0\mu m/5.8\mu m$ diagnostic plane (Fig. 11). We also show the trends to be expected for a variety of the most likely emission mechanisms.

In the first place, many PNe show evidence for shock or fluorescently excited $H_2$ emission, which can be particularly important in the IRAC bands (e.g. Phillips, Ramos-Larios & Guerrero 2011). The S(4)-S(23) v = 0-0 transitions extend over this entire region, whilst the v = 1-0 O(5)-O(8), v= 2-1 O(5)-O(7), and v=3-2 O(4) and O(5) transitions are relevant in the two shorter wave IRAC channels.

For a Boltzmann distribution of states, the column densities relative to the v = 0, J = 4 state are given through

$$\frac{g_4 N(v,J)}{g_J N(0,4)} = \exp\left\{-\frac{E(v,J) - E(0,4)}{kT_{ex}}\right\} \ldots\ldots(1)$$

where the left-hand side of the equation is also equal to

$$\frac{F(v',J')\nu_{0,2S(2)}A_{0,4\to 0,2}g_4}{F(0,4)\nu_{\Delta v,\Delta J}A_{v',J'\to v'',J''}g_J} \ldots\ldots(2)$$

where $T_{EX}(J)$ is the excitation temperature of the rotational levels, $F(v',J')$ is the observed line flux at frequency $\nu_{\Delta v,\Delta J}$, and $A_{v',J'\to v'',J''}$ is the transition probability. We have used these equations, and Einstein A coefficients from Turner et al. (1977) (see also the summary by Darren L. DePoy at http://www.astronomy.ohio-state.edu/~depoy/research /observing/molhyd.htm), to determine the IRAC band ratios illustrated in Fig. 11 , where trends are shown for the 3.6μm/5.8μm ratios (lower dashed line) and 4.5μm/5.8μm ratios (upper line). We have assumed excitation temperatures ranging between 750 K to 3000 K, and that excitation temperatures of the rotational and vibrational states are comparable.

It is clear that where 3.6μm/5.8μm ratios are similar to those which are observed (indicated by the large filled circle to the upper right-hand side), then 4.5μm/5.8μm ratios would be comparable or very much greater – in contrast to the trends in Fig. 10, where it is evident that 4.5 μm fluxes are small. It follows that observed fluxes are inexplicable in terms of shock excited $H_2$ emission alone – although it is possible that UV fluorescence would lead to markedly differing results.

It has been noted, apart from this, that MIR fluxes may be influenced by PAH emission bands – and are often found to dominate fluxes in the IRAC photometric channels. An indication of the corresponding ranges in 3.6μm/8.0μm and 8.0μm/5.8μm ratios may be obtained from the strengths of the 3.3, 6.2, 7.7 and 8.6 μm PAH emission bands.

The ratios 3.3μm/6.2μm for Galactic sources typically range between 0.03 and 0.5 (e.g. Hony et al. 2001), whilst the larger part of the 8.6μm/7.7μm and 6.2μm/7.7μm band ratios are concentrated in the ranges ~ 0.1-0.5 and ~ 0.2-0.6 (Chan et al. 2001). We have used these values to determine the region of the PAHs within Fig. 11, indicated using the rectangular box to the right-hand side.

It is clear, from this, that the PAH emission bands are much more successful in representing our ratio results – although the SW LISp still lies outside of the PAH emission region. Either one must assume that the PAHs in NGC 2371 are somewhat unusual, or that some other contribution is modifying the results.
We finally note that a variety of transitions can also influence these results. Where shocks result in enhanced ionic emission, for instance, then Reach et al. (2006) find that emission in IRAC bands 1, 3 & 4 (i.e. 3.6 μm, 5.8 μm, & 8.0 μm) have relative values 0.01/0.74/1.0. This leads to a vectorial displacement to the lower right of Fig. 11, as indicated by the arrow on the left-hand side.

So, we have seen that neither shocked $H_2$ transitions nor PAH emission bands are capable, on their own, of explaining the present LISp results. It is however possible that PAH emission components combined with HI transitions would permit some concordance with our present results. This would likely be aided by the contribution of the $\lambda 3.092$ μm transition of He II, and perhaps by low excitation transitions in the 8.0 μm channel.

If this is the case, then it is relevant to ask why PAH emission should be so strong in these regions. We have already noted that PAHs may explain the trends in MIR ratio noted in the profiles in Fig. 4 & 5, and in the ratio mapping in Fig. 6 (see Sect. 4). Why should such bands be particularly strong within the LISp emission regions, however?

One possibility is that the LISp responsible for creating the neutral condensations hypothesized in Sect. 6.1 have also punched a "hole" within the primary envelope of the source, permitting FUV photons to escape to larger radii, and leading to enhanced pumping of the PAH molecules. It is also conceivable that shock interaction of the LISp with neutral gas leads to the shattering of larger grains, in a manner similar to that described by Allain, Leach & Sedlmayr (1996) and Jones,

Tielens & Hollenbach (1996). This would result in enhanced volume densities of smaller grain particles, and higher levels of excitation for the associated PAHs.

## 7. Conclusions

We have processed extensive archival material for the planetary nebula NGC 2371, including narrow band imaging acquired with the JKT and *HST*, and MIR images obtained using *Spitzer*. These have been used to describe the structure of the source, and evaluate likely emission mechanisms.

These observations also show that the interior structure of the shell is relatively complex, and that [OIII] emission is strong at the limits of the envelope, where it takes the form of an unresolved bright emission rim. This structure is similar to the rims observed in several other PNe, where they have been attributed to local shock activity. It is possible that the outer rim of NGC 2371 lies at the interface with the AGB mass-loss halo.

*HST* imaging of the source in [NII], H$\alpha$ and [OIII] again shows evidence for the [OIII] rims noted in the JKT results. By far the most interesting aspect of the latter observations, however, is the presence of narrowly defined low-ionization spokes (LISp) located along a PA of ~65° – an orientation which is at variance with the PAs of the lobes, or the major and minor axes of the inner shell. It is possible that such "off-axis" LISp arise through precession of the central star. The LISp also appears to be associated with [OIII] emission collars, giving the impression that the LISp may have punched their way though the nebular shell. Although the nature of these collars still remains uncertain, the sharply defined rims of these structures suggests ongoing shock activity, or formation in the recent past.

It is clear that the LISp are composed of many low-excitation striations, and we suggest that these correspond to UV shadow zones, or material which has been ram-pressure stripped from highly compact (< 1 arcsec) condensations. Such hypotheses are consistent with relative line ratios in this region.

The LISp, collars and lobes are also visible in the MIR, where we note that the overall morphology of the shell is similar to that observed in the visible. It is pointed out that the strongest emission occurs at 4.5 $\mu$m, at variance with the tendencies observed in other PNe, whilst there is evidence that 8.0$\mu$m/4.5$\mu$m and 5.8$\mu$m/4.5$\mu$m flux ratios increase to larger radii, perhaps achieving their largest values at the limits of the lobes. Such trends, which are also observed in other PNe, are often attributed to emission by PAHs in external PDRs.

Although the MIR emission characteristics of the LISp are difficult to define, it seems likely that the SW LISp has ratios 3.6$\mu$m/5.8$\mu$m $\cong$ 1.85, and 8.0$\mu$m/5.8$\mu$m $\cong$ 5.44. Levels of emission at 4.5 $\mu$m are likely to be negligibly small. If this is accepted, then the results can be explained in terms of enhanced levels of PAH emission, together a variety of permitted and forbidden line transitions. The enhanced PAH emission may arise where dust grains are shattered in local shocks, leading to increased volume densities of smaller grains, or where levels of FUV emission are greater than in adjacent portions of the shell, leading to enhanced FUV pumping of the PAHs.

**Acknowledgements**

GRL acknowledges support from CONACyT and PROMEP (Mexico).

Also we want thank to an anonymous referee who made very excellent comments and suggestions for the improvement of the paper.

We would like to dedicate this paper in memory of our colleague and friend Prof. John Peter Phillips who recently passed away.

# Figure Captions

**Figure 1**

Images of NGC 2371 in the MIR (upper panel) and visible (lower panel). In the former case, we have combined Spitzer imaging of the source taken at 3.6 $\mu$m (blue), 4.5 $\mu$m (green) and 8.0 $\mu$m (red). Note the evidence for weak low-ionization features in the outer portions of the shell, along a PA of ~65°, and the outer limits of the bipolar structure along a PA of ~120°. In the lower panel, we show an [OIII] $\lambda$5007 image of the source taken with the JKT, processed so as to reveal finer details of the nebular structure. We note evidence for an enhanced rim of emission at the outer limits of the shell, as well as a complex bilobal structure along the major axis of the source.

**Figure 2**

Images of NGC 2371 taken with the HST. The left-hand panel shows a combined exposure in [NII] $\lambda$6584 (red), [OIII] $\lambda$5007 (blue) and H$\alpha$ (green). Note the evidence for small, compact low-excitation clumps throughout the nebular shell, and the low-ionization spokes along PA ~65°. The right hand panel shows the corresponding ratio of [OIII]/H$\alpha$. Although these ratios are for most part relatively uniform, we see evidence for barely resolved rims of enhanced [OIII] emission, and strongly reduced [OIII]/H$\alpha$ ratios within the LISp region. The sudden rise of [OIII]/H$\alpha$ around the zones of the two LISp is probably the result of a smooth gradient in the fractional [OIII] abundances in their vicinity. Even the changes near the LISp of this ratio will suggest that the ionization shadows are cast by material in the zones where the RGB Spitzer image (Fig. 1) is yellow in colour.

**Figure 3**

Spitzer MIR and HST visual profiles along the LISp axis of NGC 2371, where we note evidence for strong deviations in the profiles associated with the LISp, and with their associated collars of [OIII] emission (referred to as the NE and SW rims). The MIR profiles are indicated in units of MJy sr$^{-1}$, whilst the units of the HST profiles are arbitrary, although relative line strengths are accurately portrayed. The directions and widths of the profiles are indicated in the inserted figure.

**Figure 4**

The variation of line and flux ratios along the LISp axis of NGC 2371, where HST results are illustrated in the upper panel, and Spitzer profiles are in the lower panel. Note the strong deviation in the profiles associated with the LISp and shock regions, particularly evident in the SW portions of the shell. The direction and widths of the profiles are indicated in the inserted figures, which correspond to an [NII] image (in the top panel), and an 8.0 $\mu$m image (in the lower panel).

**Figure 5**

MIR profiles along the major axis of NGC 2371 (top panel), and the corresponding variation in flux ratios (bottom panel). It will be noted that 4.5 $\mu$m emission is strongest in the centre of the source, although 8.0 $\mu$m emission is dominant in the ansae. There is also evidence for an increase in ratios to larger radial distances (i.e. for RPs > 15 arcsec).

**Figure 6**

MIR flux ratio mapping for NGC 2371, where the bars at the top of the panels indicate the grey-scale calibrations. All of the maps have similar structures, and suggest lower ratios at the centre of the source, increasing to larger values at the periphery.

**Figure 7**

The variation of ionic line ratios through a condensation at the base of the SW LISp (upper panel), and across the nearby [OIII] collar (lower panel). The directions and widths of the profiles are indicated in the inserted panels. Ratios have been normalised to unity at RP = -0.3 arcsec in the upper panel, and RP = 0 arcsec in the lower panel. This permits a clearer comparison of the differing line ratio trends. It is evident that the [NII]/H$\alpha$ and [SII]/H$\alpha$ ratios increase markedly as one enters the LISp region, although [OIII]/H$\alpha$ ratios decrease. By contrast, there seems to be little variation in lower excitation ratios close to the rim of the [OIII] collar (lower panel), even though [OIII]/H$\alpha$ ratios decline quite markedly.

**Figure 8**

A montage showing a combination of [OIII] (blue), Hα (green) and [NII] (red) images of the SW LISp (upper left-hand panel), and the ratios of [OIII], [NII] and [SII] with Hα (where higher ratios correspond to darker shades of grey). It is clear that the LISp is composed of multiple tails, extending radially behind highly compact condensations. It is also evident that the condensations/tails possess higher levels of low excitation emission.

**Figure 9**

The distribution of HST line ratios in the SW LISp, where LISp values are indicated with small green diamonds; Galactic PNe using larger blue diamonds; and the shock modelling results of Hartigan et al.(1987) and Shull & McKee (1979) are indicated using red disks. The regions for SNRs, HII regions and PNe are taken from Sabbadin, Minello, & Bianchini (1977) and Riesgo & Lopez (2006). It seems clear that ratios for the LISp are consistent with photoionised emission, and displaced from the shock modelling results (upper panel) and of the region for SNRs (lower panel).

**Figure 10**

The variation in surface brightnesses close to the SW LISp, where the width and direction of the traverse is indicated in the inserted panel. The results are normalised to unity at RP = 14.65 arcsec. Note how there is a strong deviation in the 3.6, 5.8 and 8.0 μm results close to the positions of the LISp, whilst the trend at 4.5 μm appears to be relatively unaffected. We interpret this as indicating that levels of 4.5 μm emission are low within the LISp.

**Figure 11**

The location of the SW LISp within the 4.5μm/5.8μm-8.0μm/5.8μm colour plane, where we also indicate trends to be expected for a variety of emission mechanisms. Thus, the PAH region is defined using the results of Chan et al. (2001) and Hony et al. (2001), whilst the vectorial trends on the left-hand side (i.e. those for warm dust, shocked and low excitation ions, and HI emission) are based on an investigation of ISO spectra, and work presented by Reach et al. (2006) and Osterbrock (1989). The variation of ratios due to the v = 0-0 transitions of $H_2$ are determined for excitation temperatures $T_{EX}$ = 750-3000 K; where the upper dashed line indicates the variation of 4.5μm/5.8μm ratios, whilst the lower trend is for 3.6μm/5.8μm ratios.

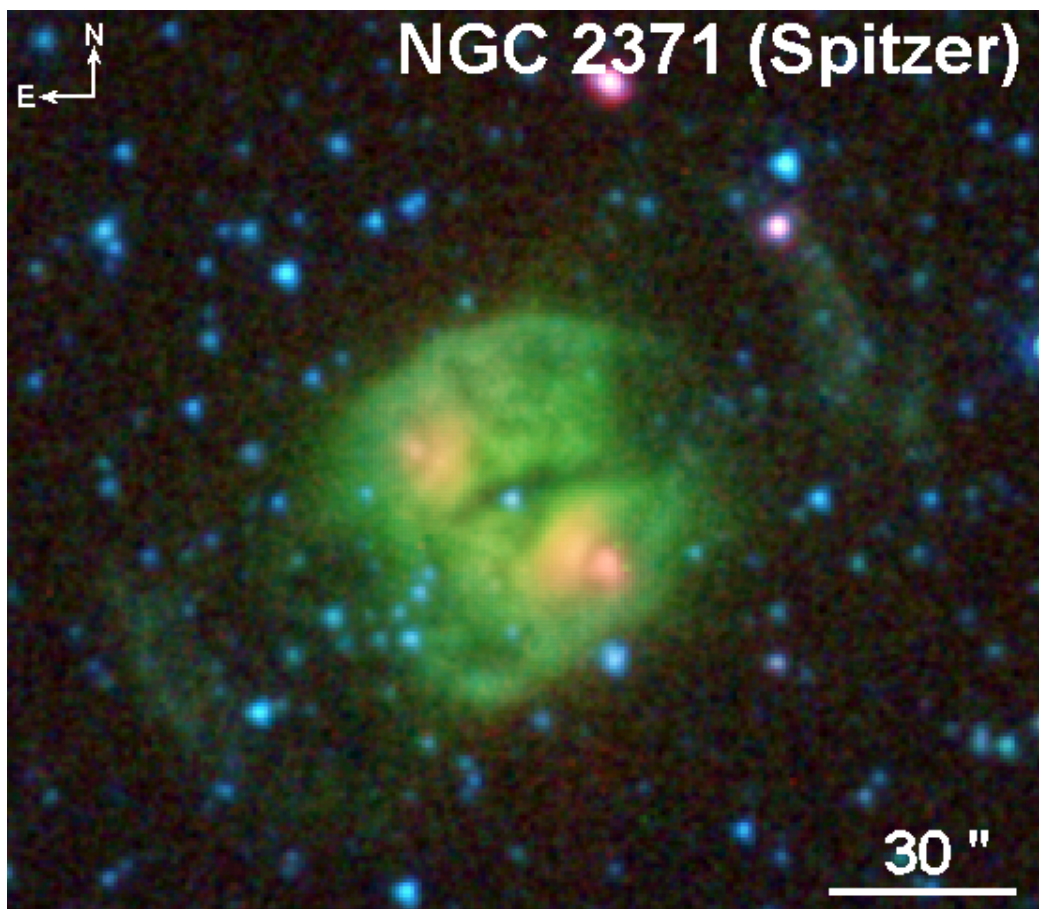
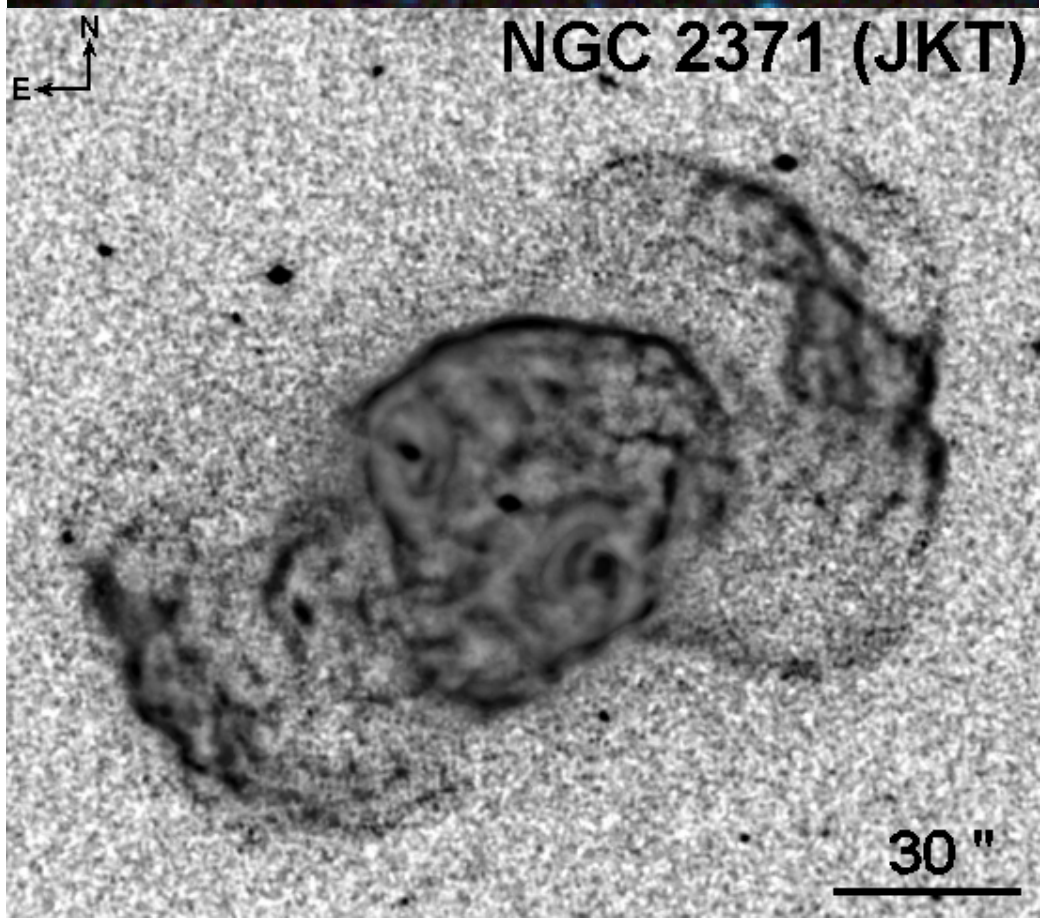

FIGURE 1

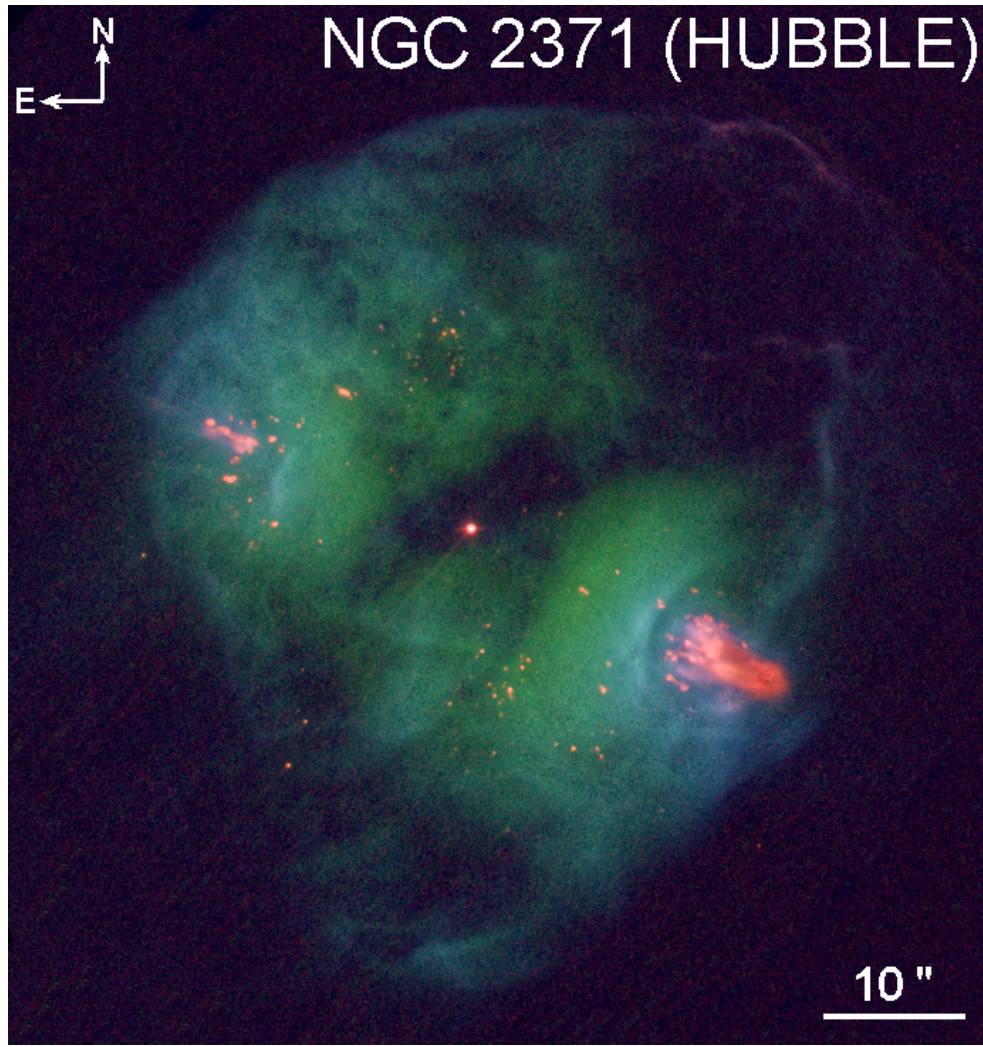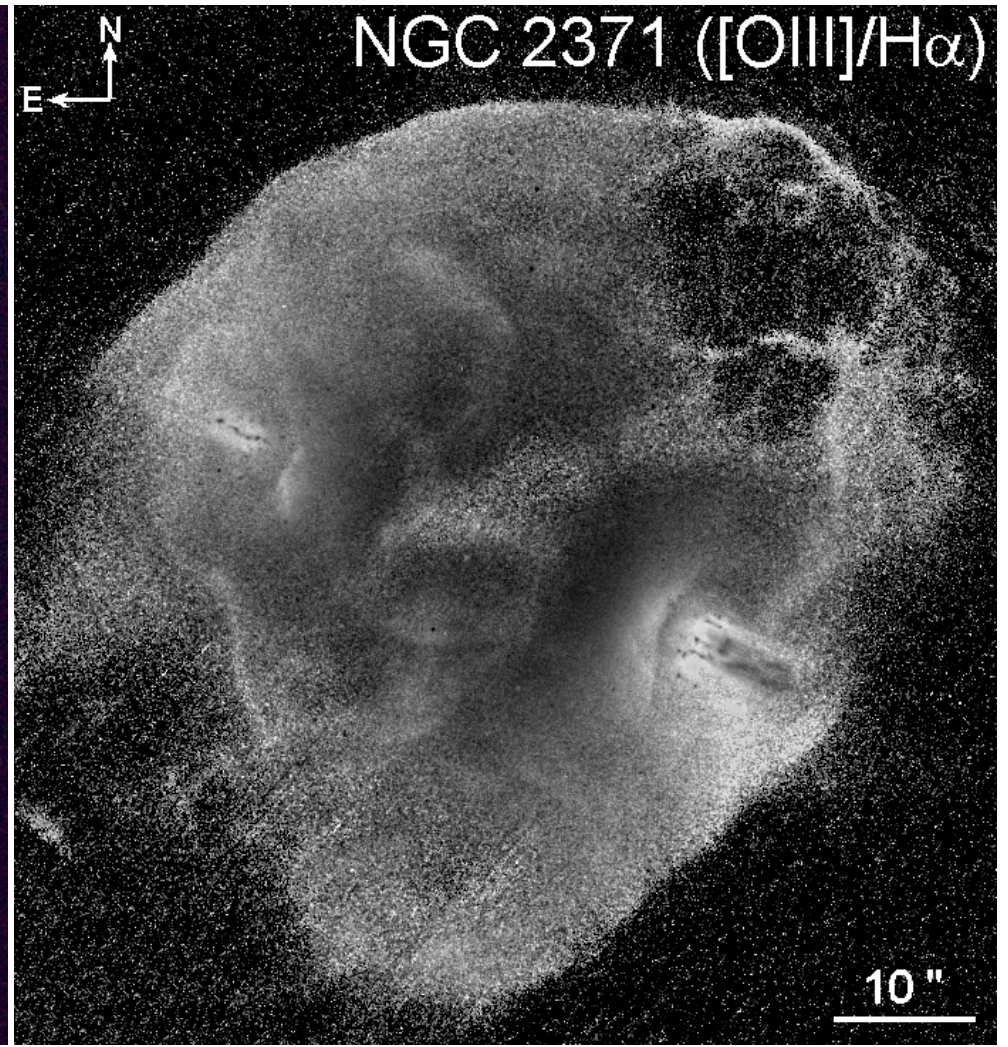

FIGURE 2

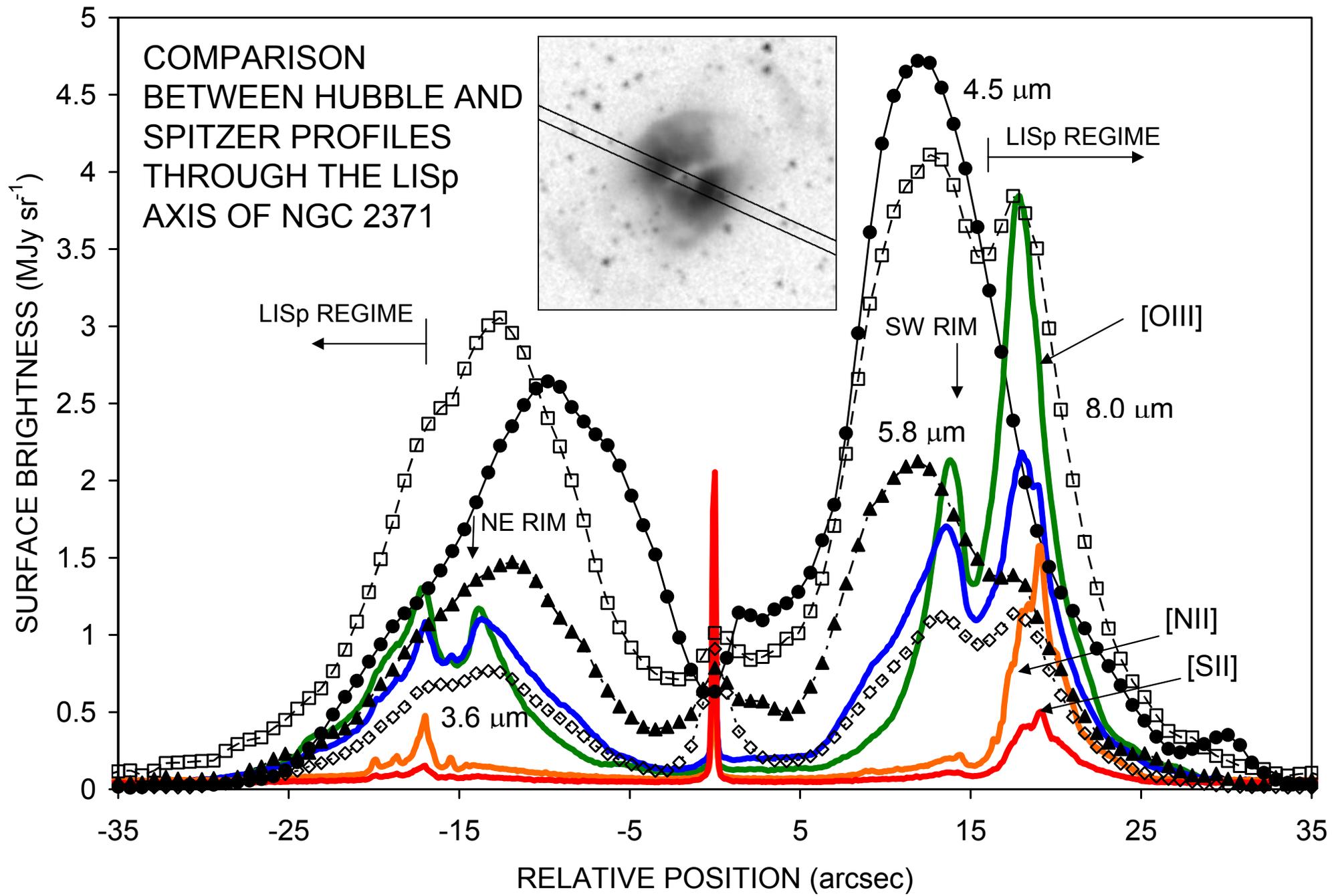

FIGURE 3

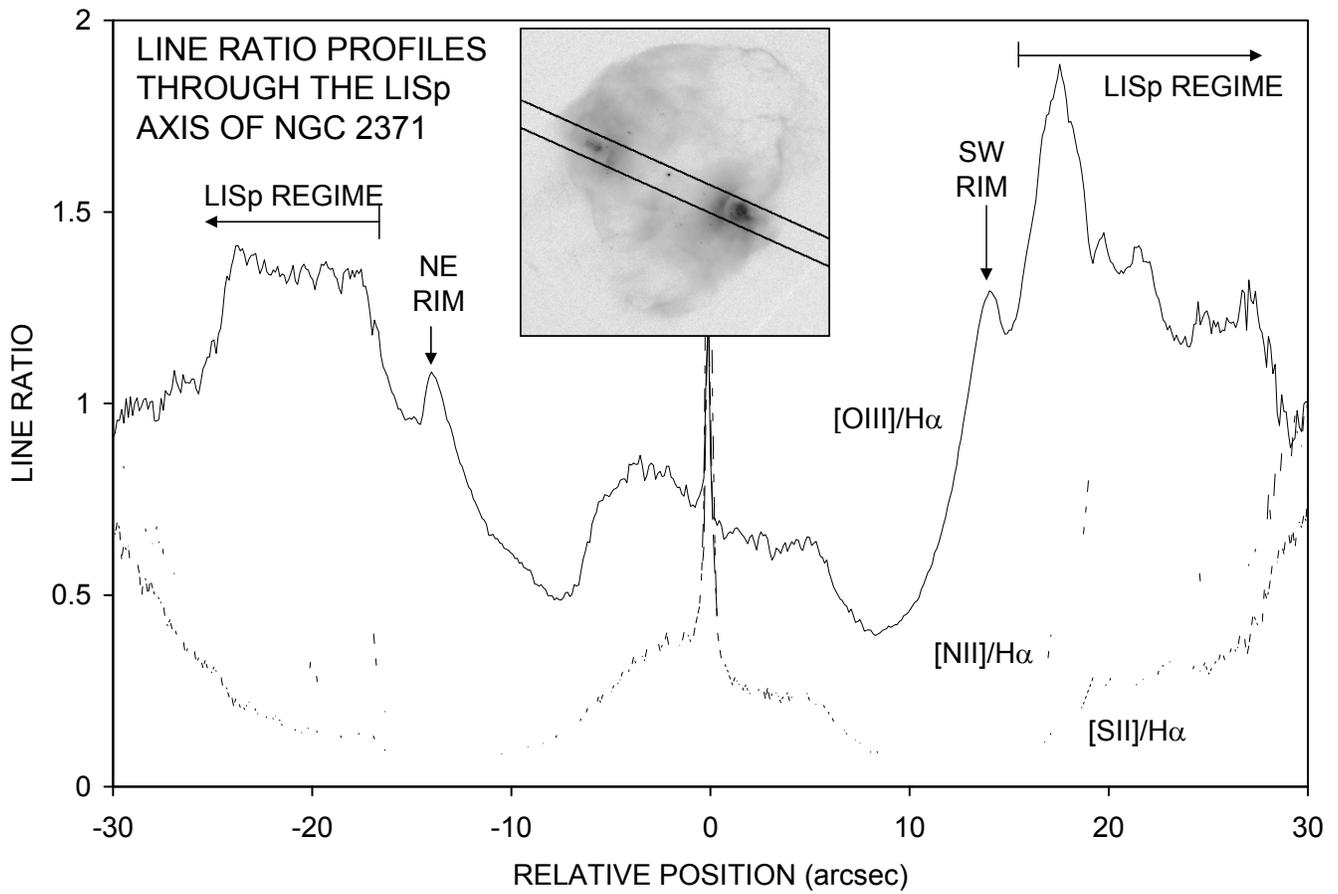
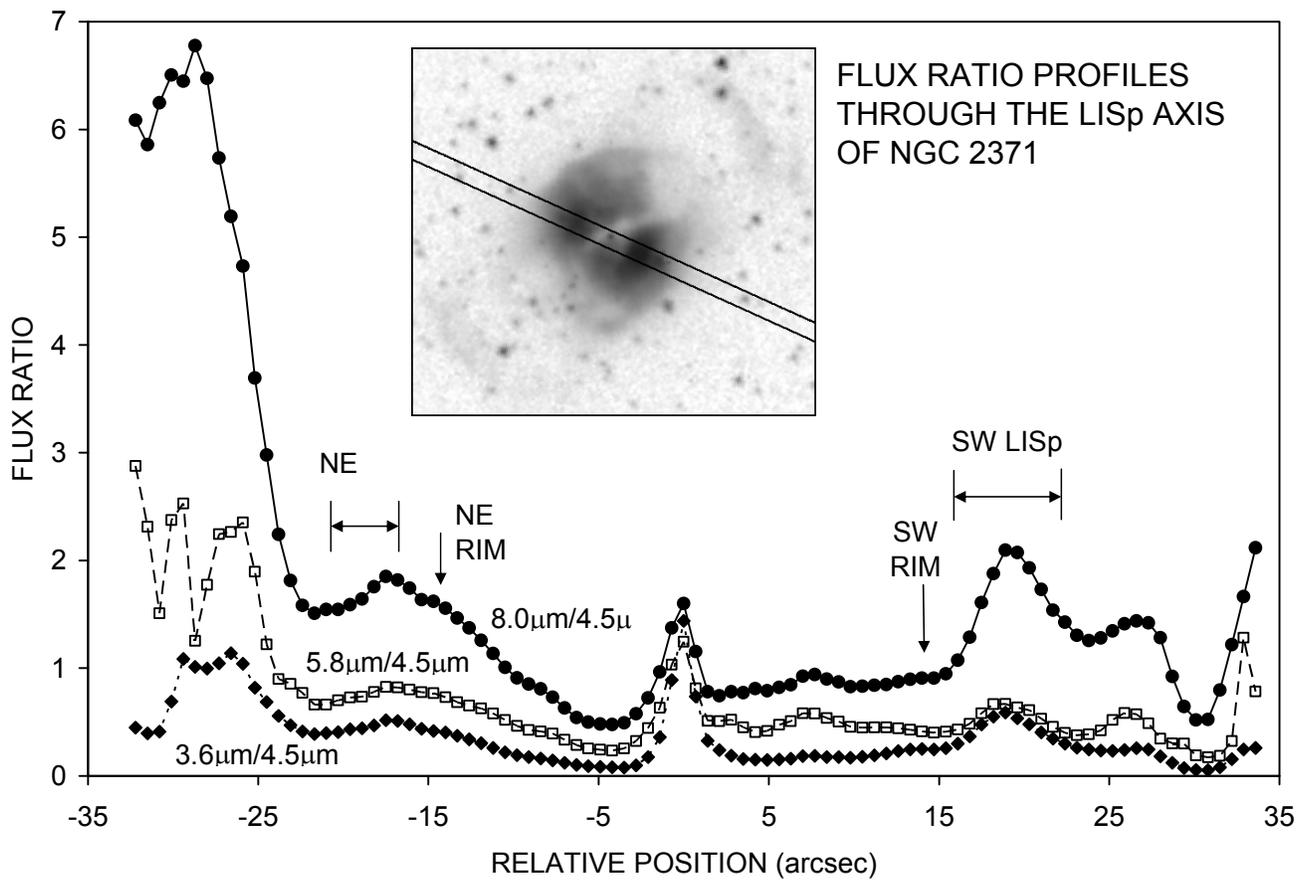

FIGURE 4

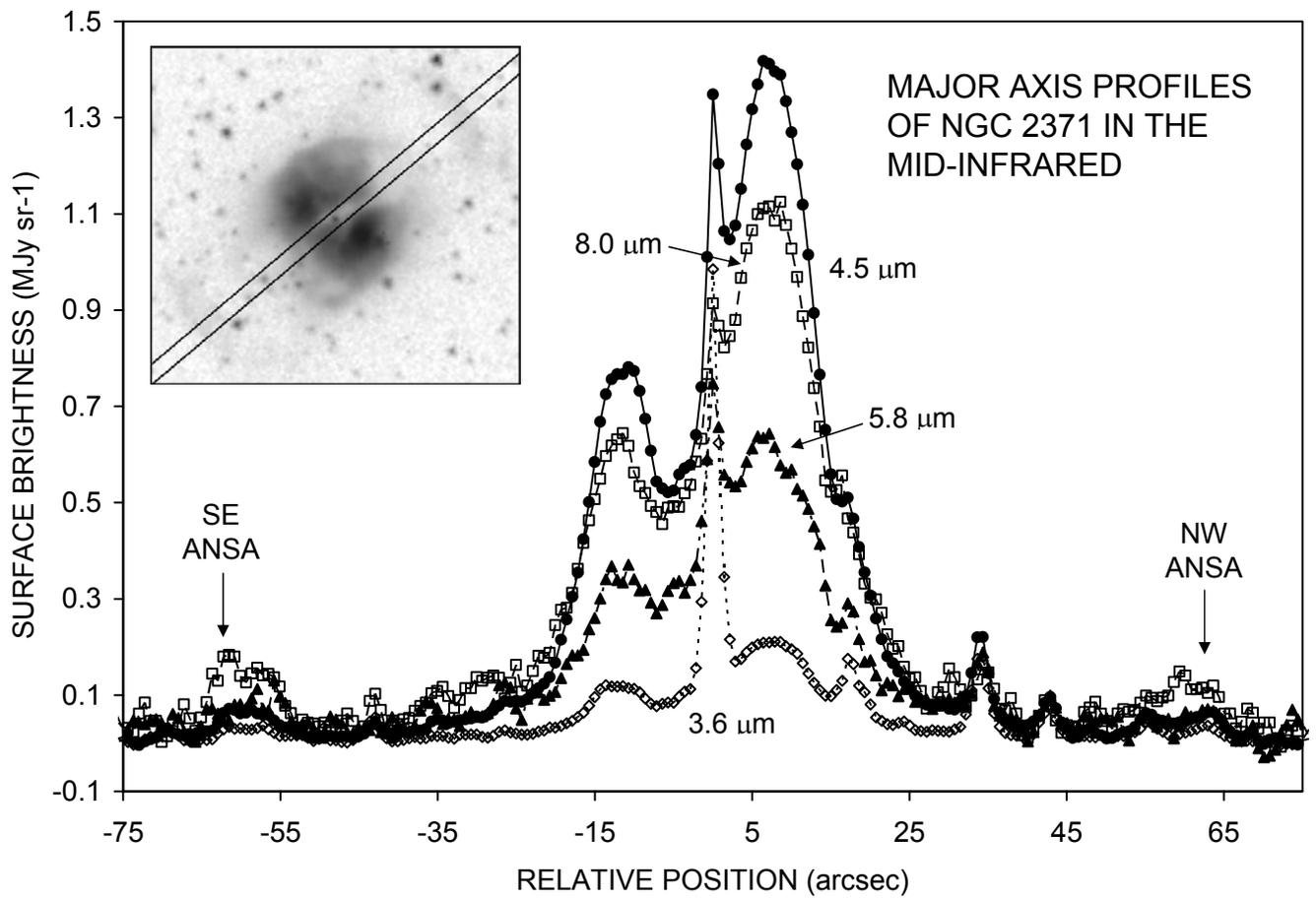
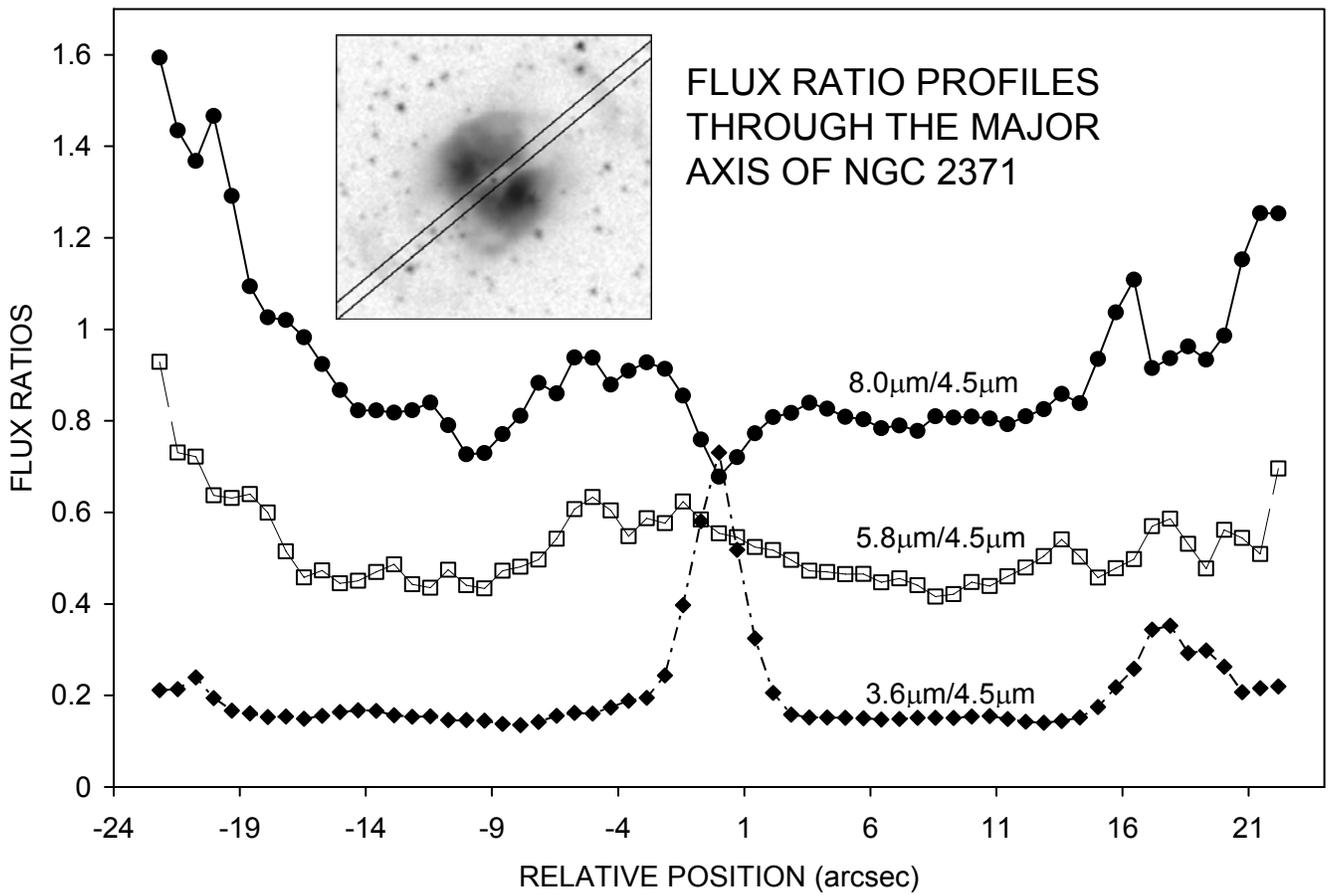

FIGURE 5

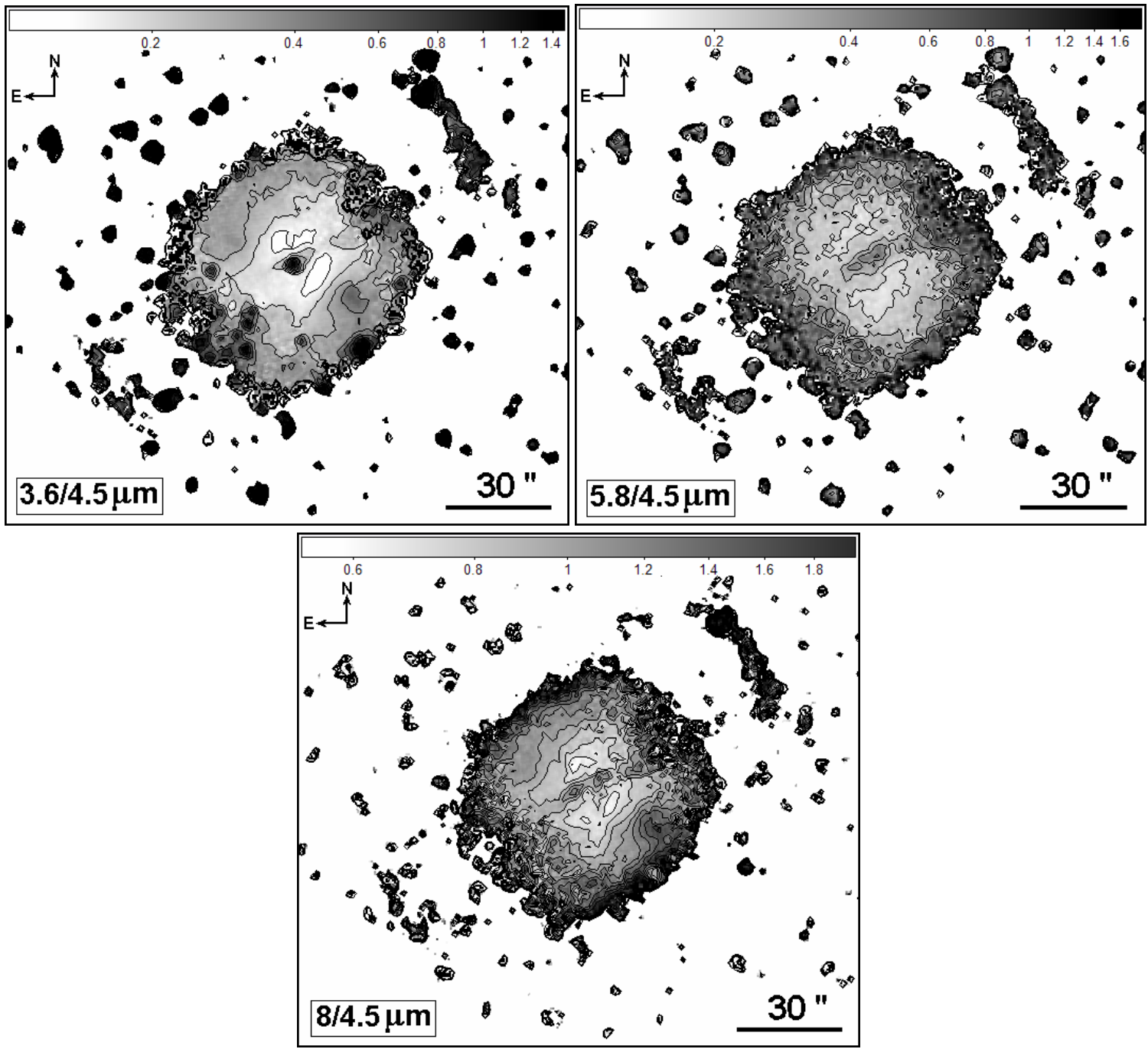

FIGURE 6

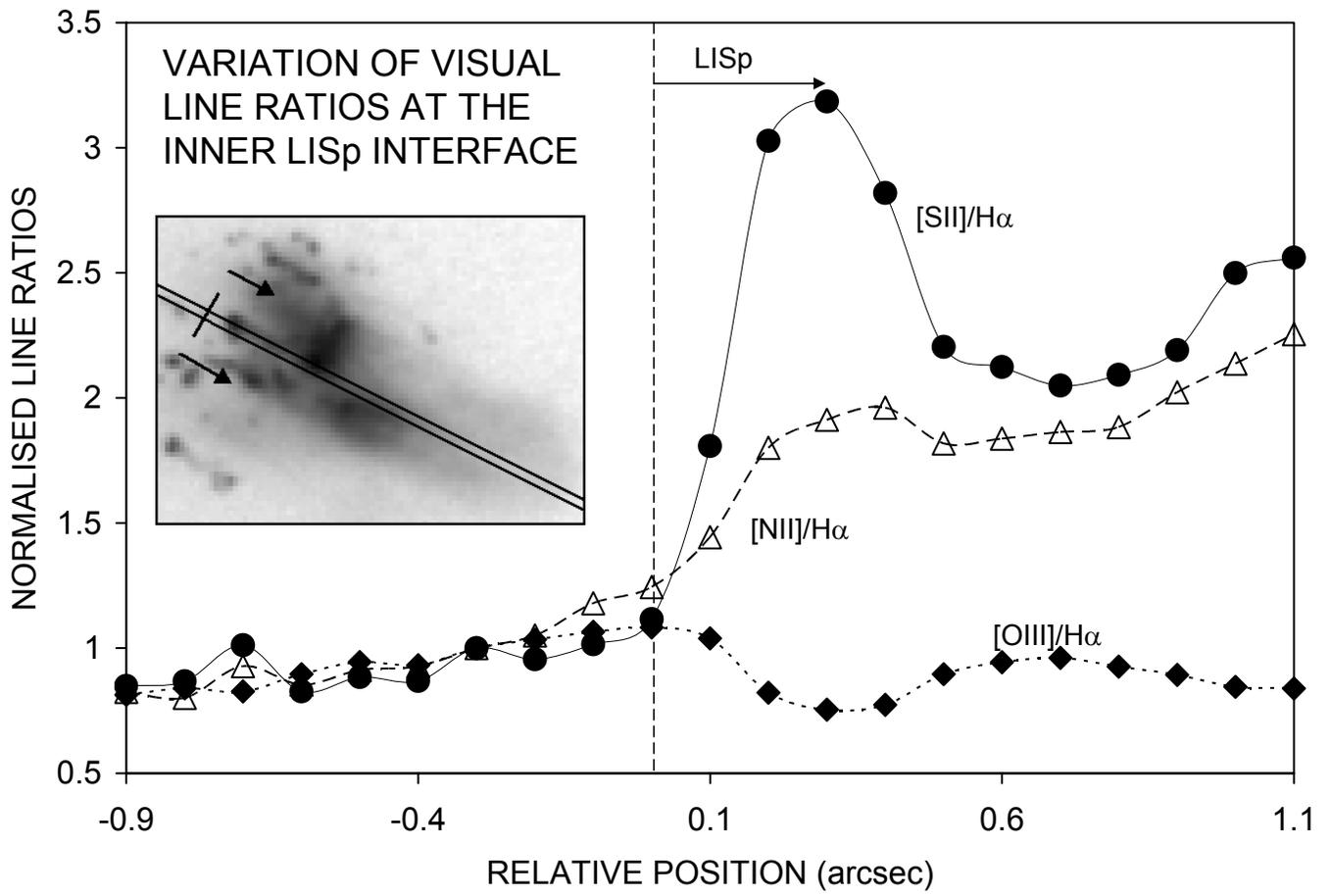

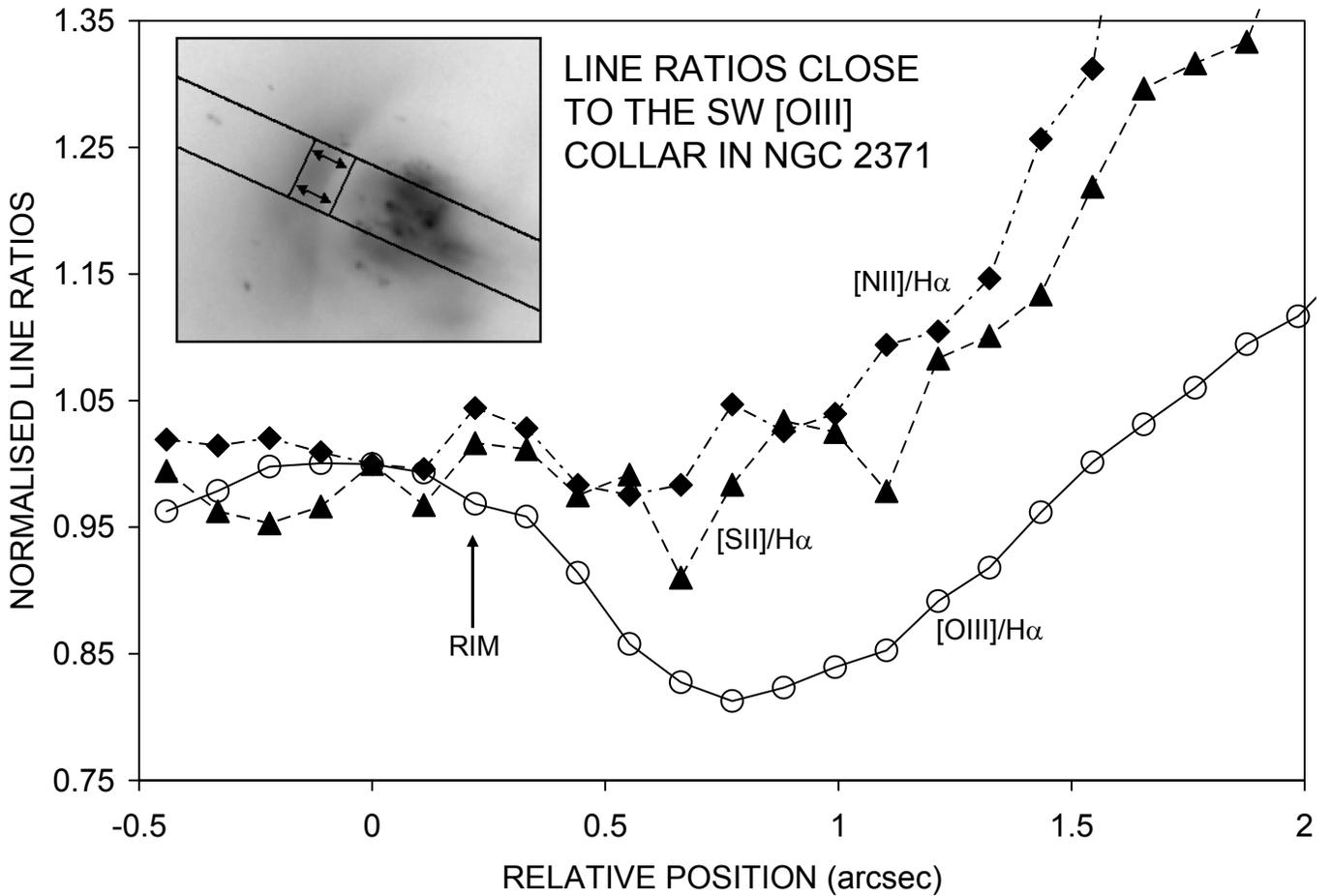

FIGURE 7

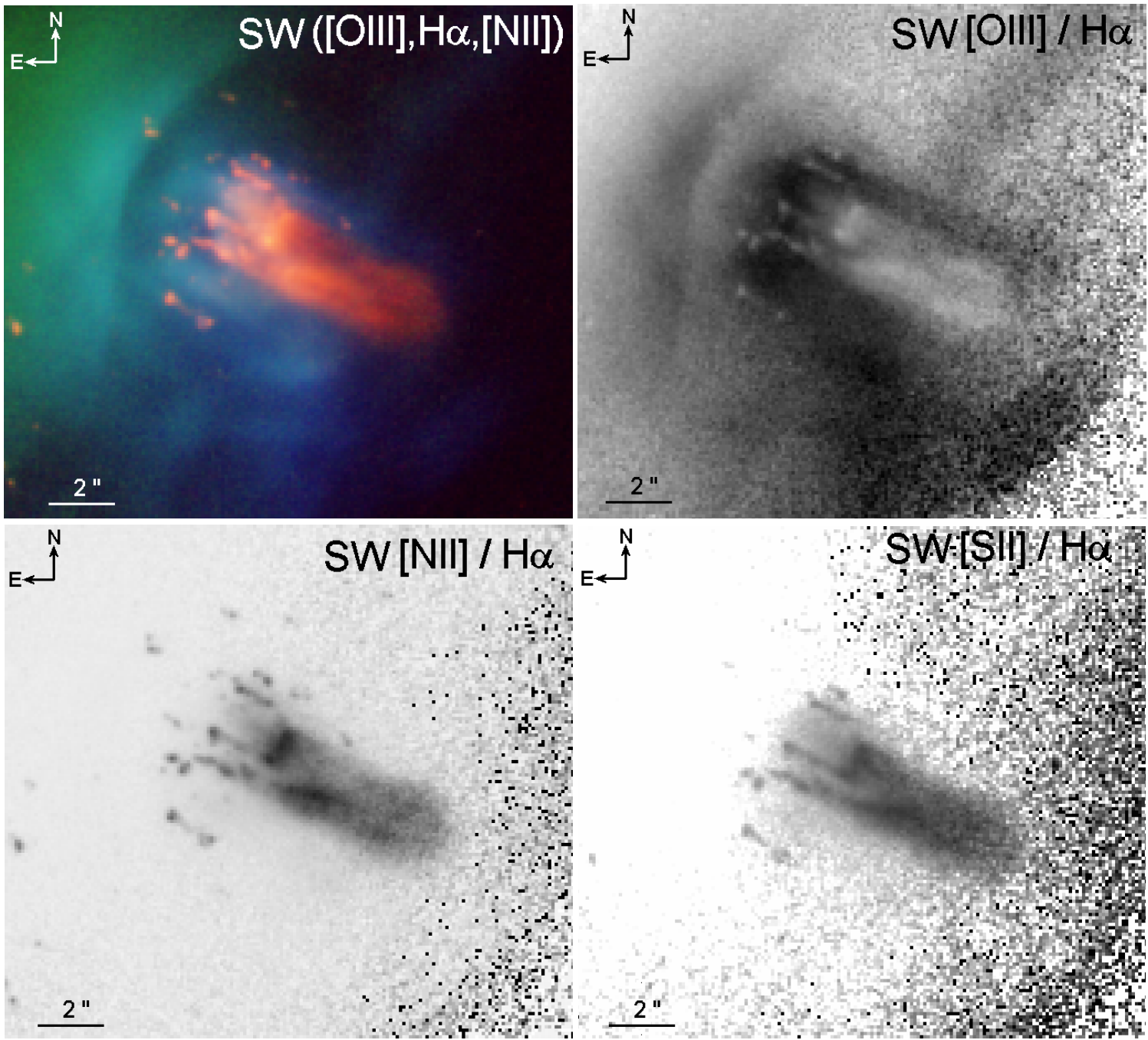

FIGURE 8

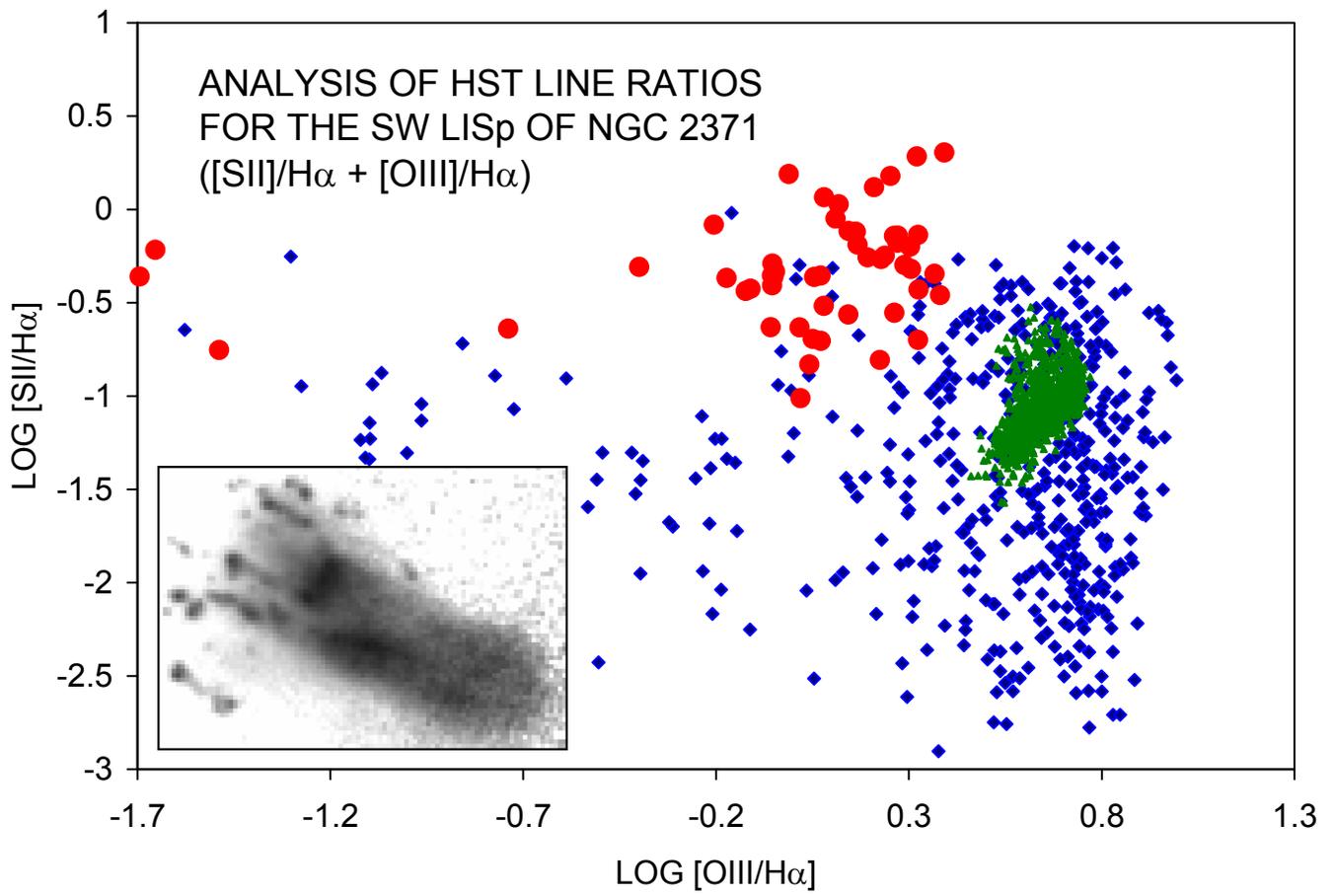

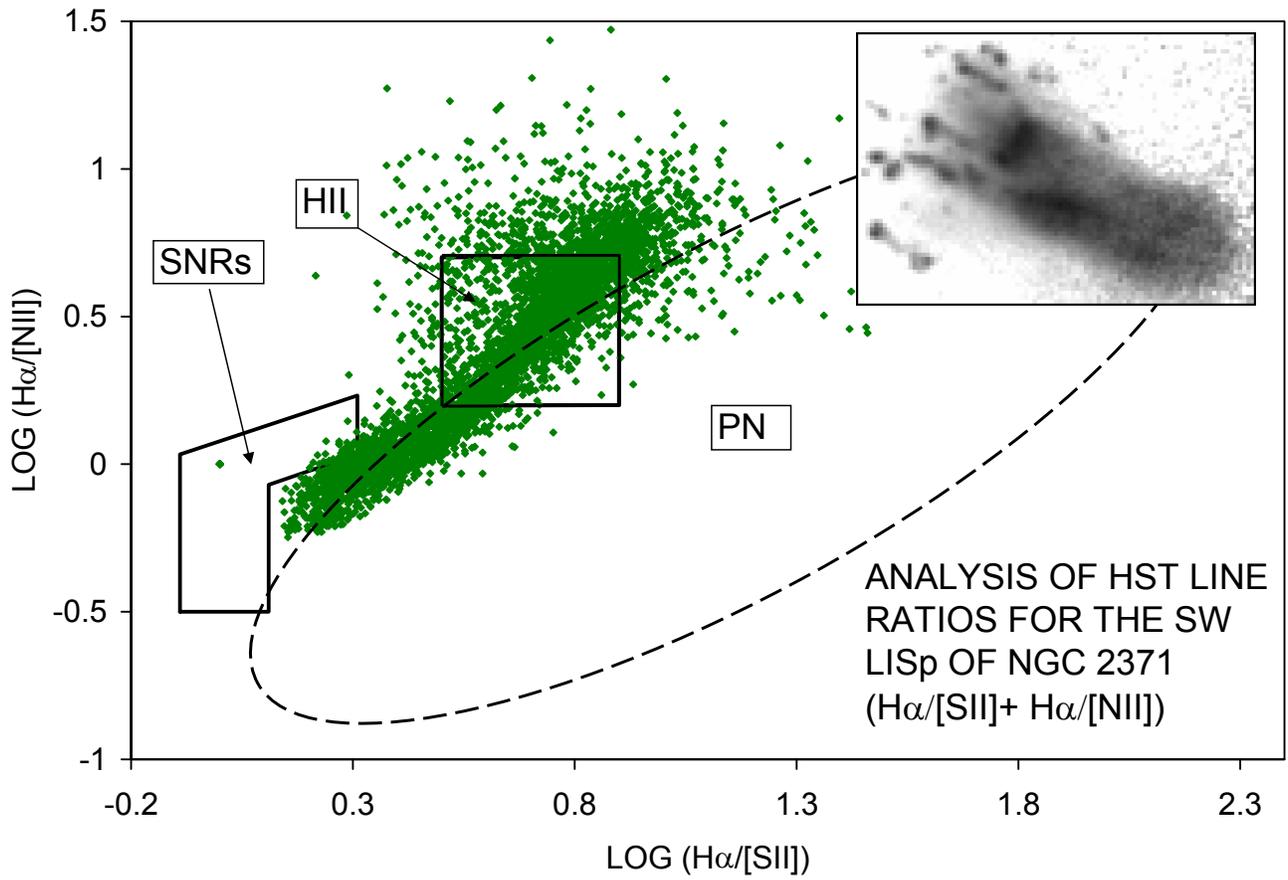

FIGURE 9

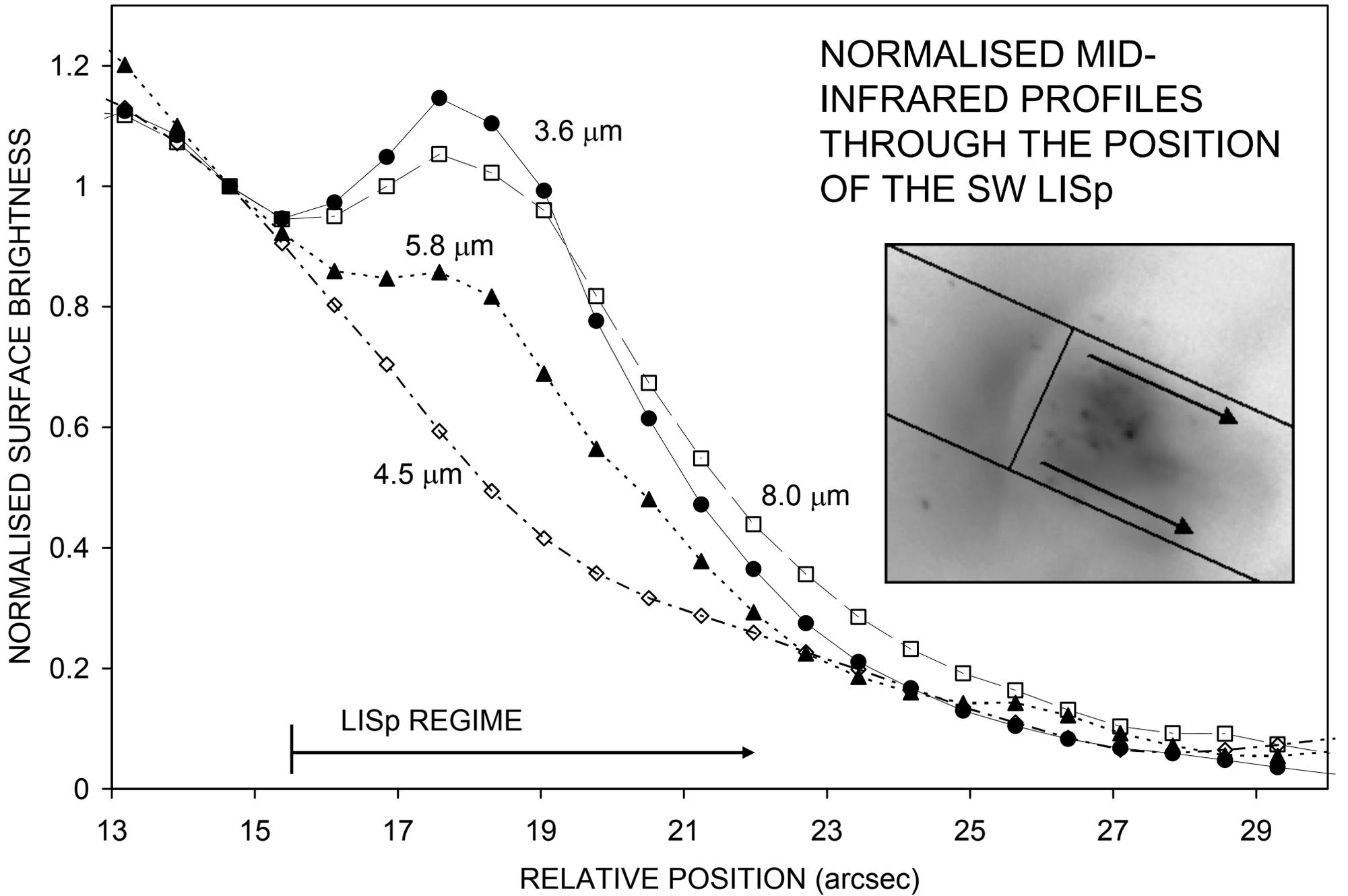

FIGURE 10

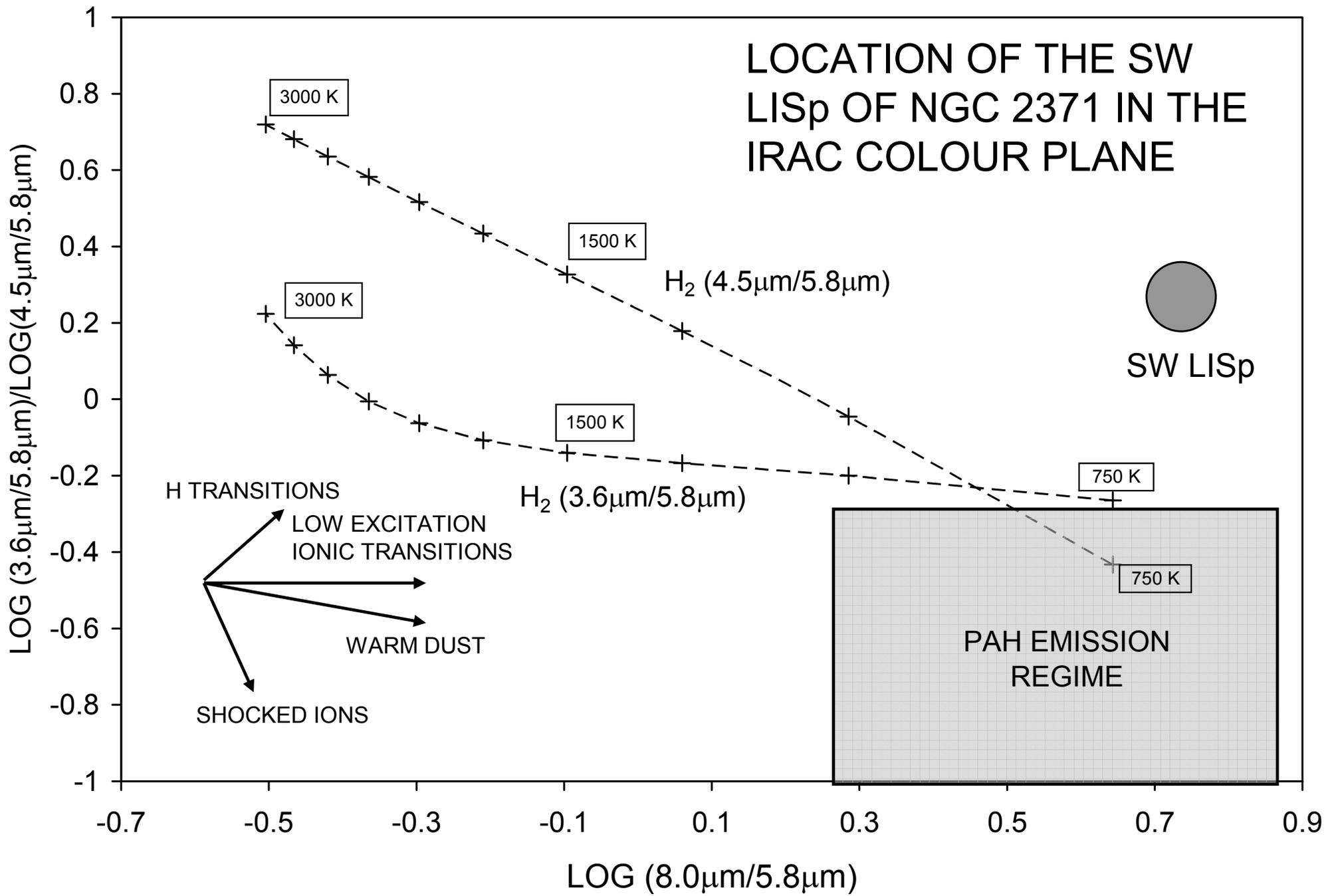

FIGURE 11